# Dipolar interaction effects in the thermally activated magnetic relaxation of two-dimensional nanoparticle ensembles

S. I. Denisov[a)] and T. V. Lyutyy

*Department of Mechanics and Mathematics, Sumy State University, 2, Rimskiy-Korsakov Street, 40007 Sumy, Ukraine*

K. N. Trohidou[b)]

*Institute of Materials Science, NCSR "Demokritos", 15310 Athens, Greece*

**Abstract**

The thermally activated magnetic relaxation in two-dimensional ensembles of dipolar interacting nanoparticles with large uniaxial perpendicular anisotropy is studied by a numerical method and within the mean-field approximation. The role that the correlation effects in the presence of a bias magnetic field and taking into account the lattice structure play in magnetic relaxation is revealed.

The two-dimensional (2D) ensembles of uniaxial ferromagnetic nanoparticles, whose easy axes of magnetization are perpendicular to the nanoparticles plane, represent an important class of perpendicular magnetic recording media.[1] From the technological point of view, the main characteristic of these media is the average time of data storage, which is one of the numerical parameters that describe the thermally activated magnetic relaxation. Due to the dipolar interaction between nanoparticles, the analytical description of magnetic relaxation in such ensembles is a very complicated problem, which was solved only in the case that the mean[2,3] and fluctuating[4] components of the dipolar field were taken into account. However, the correlations of the nanoparticle magnetic moments, arising from the dipolar interaction, also play an important role, especially on the final phase of the magnetic relaxation. To study the features of magnetic relaxation, which are conditioned by the correlation effects, we have developed a method for its numerical simulation at zero bias field.[5] At present only this method permits the calculation of the relaxation law on times exceeding the largest relaxation time; the time-quantified[6] and conventional[7,8] Monte Carlo methods are not suitable for this purpose, since the former is applicable on substantially smaller times,[5] and the latter has no physical time associated with each Monte Carlo step.

In this letter, we extend our method[5] in the case of nonzero bias fields, and we present results, obtained within its framework, which clarify the role that the correlation effects, the bias field and lattice structure play in the magnetic relaxation. We consider the 2D ensembles of spherical nanoparticles with a radius $r$, which occupy the sites of a square or hexagonal lattice with a lattice spacing $d$, and whose easy axes of magnetization are perpendicular to the lattice plane ($xy$ plane). We also assume that the nanoparticle magnetic moments $\mathbf{m}_i(t)$ (the index $i$ labels the nanoparticles) perform a coherent rotation ($|\mathbf{m}_i(t)| = m$), on each $\mathbf{m}_i(t)$ acts a bias magnetic field $H_0\mathbf{k}$ ($\mathbf{k}$ is the unit vector along the $z$ axis) and that the conditions $\mathbf{m}_i(0) = m\mathbf{k}$ hold at the initial time $t = 0$.

If the smallest heights $\Delta U_i$ of the potential barriers between the equilibrium directions of all $\mathbf{m}_i(t)$ are much larger than the thermal energy $k_B T$ ($k_B$ is the Boltzmann constant, $T$ is the absolute temperature), i.e., $e_i = \Delta U_i / k_B T \gg 1$ for all nanoparticles, then the vectors $\mathbf{m}_i(t)$ fluctuate within small vicinities of the $\mathbf{k}$ and $-\mathbf{k}$ directions, and they are reoriented occasionally. In this case the average numbers $N_+(t)$ and $N_+(t)$ of positively and negatively oriented magnetic moments in a lattice region that contains $N (\gg 1)$ nanoparticles are well-defined ($N_+(t) + N_-(t) \approx N$), and we can introduce the reduced magnetization of the nanoparticle ensemble as $r(t) = 2N_+(t)/N - 1$.

Using the approaches described in Ref. [5], we can calculate the relaxation law $r(t)$ in a wide time interval. The main advantage of this approach is that the real dipolar fields acting on the nanoparticle magnetic moments are taken into account. Within the mean-field approximation, which ignores the correlations effects, the numerical procedure is needless, and we can easily determine the mean-field relaxation law[3] $r_{mf}(t)$. We have calculated $r(t)$ and $r_{mf}(t)$ for comparison. We choose 2D ensembles of Co nanoparticles characterized by the parameters $H_a = 6400$ Oe, $m/V = 1400$ G ($V$ is the nanoparticle volume), $l = 0.2$, $r = 4$ nm, and $T = 300$ K. Depending on the lattice structure, we place the basic nanoparticle ensemble for our numerical the simulation, on square or regular hexagonal lattice with size $100d$ or $60d$, respectively. To exclude boundary effects, we assume that the basic ensemble is surrounded by eight (for a square lattice) or six (for a hexagonal lattice) identical ensembles, and each nanoparticle from the basic ensemble is considered as a central one in the square or hexagonal box of the same size (i.e., $100d$ or $60d$) and interacts only with the nanoparticles which belong to this box. We use the superscripts $^s$ and $^h$ on $r(t)$, $r_{mf}(t)$, and $d$ to denote the square and hexagonal lattice, respectively.

The role that the correlation effects and a bias field $H_0$ play in magnetic relaxation is illustrated in Fig. 1. Due to the correlations of the nanoparticle magnetic moments, for $H_0 = 0$ the actual magnetic relaxation occurs faster on small times and more slowly on large times, than the mean-field theory predicts [see Fig. 1(a)]. In this case the curves $r^s(t)$ and $r^s_{mf}(t)$ are intersected at the time $t = t_{in}$, and $r^s(\infty) = r^s_{mf}(\infty) = 0$. In equilibrium the local dipolar fields have, on average, the same directions as the magnetic moments and the mean dipolar field equals zero (for $H_0 = 0$), therefore for $H_0 \neq 0$ the condition $|r^s(\infty)| < |r^s_{mf}(\infty)|$ must hold. For $H_0 < 0$ we have $r^s_{mf}(\infty) < r^s(\infty) < 0$, and the curves $r^s(t)$ and $r^s_{mf}(t)$ are intersected only once. If $H_0 > 0$, then $r^s_{mf}(\infty) > r^s(\infty) > 0$, for small enough values of $H_0$ these curves are intersected twice [see Fig. 1(b)], with increasing $H_0$ the time interval $(t_{in}, t^*_{in})$ is decreasing, and $r^s_{mf}(t) > r^s(t)$ ($t > 0$) for large $H_0$ [see Fig. 2(a), curves 1 and 2].

The features of magnetic relaxation in ensembles of the Co nanoparticles, which arise from the different nanoparticle arrangement in square and hexagonal lattices, are demonstrated in Figs. 2 and 3. If the lattice spacing $d$ is the same for both lattices (Fig. 2), then the difference between the relaxation laws results mainly from the different number of the nearest sites in these lattices (4 vs. 6). At $t = 0$ the local dipolar fields in hexagonal lattices are larger than in square ones, so the initial phase of magnetic relaxation occurs faster in hexagonal lattices. At $t = \infty$ and $H_0 = 0$ the conditions $\boldsymbol{r}^s(\infty) = \boldsymbol{r}^h(\infty) = 0$ and $\boldsymbol{r}^s_{mf}(\infty) = \boldsymbol{r}^h_{mf}(\infty) = 0$ hold. Each magnetic moment in a square lattice is surrounded, on average, by 4 opposite oriented magnetic moments, whereas in a hexagonal lattice – by 4 opposite and by 2 similarly oriented magnetic moments, therefore, if $H_0 > 0$ [see Fig. 2(a)] then $\boldsymbol{r}^s_{mf}(\infty) > \boldsymbol{r}^h_{mf}(\infty) > 0$, $\boldsymbol{r}^h(\infty) > \boldsymbol{r}^s(\infty) > 0$, and the curves $\boldsymbol{r}^s(t)$ and $\boldsymbol{r}^h(t)$ have the unique intersection point $t = t_1$ ($t_1 \approx 13.59$ s for $H_0 = 500$ Oe). If $H_0 < 0$ [see Fig. 2(b)] then $\boldsymbol{r}^s_{mf}(\infty) < \boldsymbol{r}^h_{mf}(\infty) < 0$ and, since $\boldsymbol{r}^h(\infty) < \boldsymbol{r}^s(\infty) < 0$, the curves $\boldsymbol{r}^s(t)$ and $\boldsymbol{r}^h(t)$ have two intersection points $t = t_1$ and $t = t_2$ ($t_1 \approx 0.01$ s, $t_2 \approx 0.33$ s for $H_0 = -500$ Oe).

Figure 3 presents the character of magnetic relaxation in square and hexagonal lattices, which are characterized by the unit cells that have the same area, i.e., $(d^s)^2 = (d^h)^2 \sqrt{3}/2$, at $H_0 = 0$. Despite the fact that each site in the hexagonal lattice has more nearest sites, the local and mean dipolar fields in the square lattice exceed the corresponding fields in the hexagonal one, since $d^s < d^h$ ($d^s \approx 0.931 d^h$). Therefore the conditions $\boldsymbol{r}^h(t) > \boldsymbol{r}^s(t)$ and $\boldsymbol{r}^h_{mf}(t) > \boldsymbol{r}^s_{mf}(t)$ must hold for all $t > 0$ [at large times (see Fig. 3b) the distinction between $\boldsymbol{r}^h_{mf}(t)$ and $\boldsymbol{r}^s_{mf}(t)$ is not visible on the chosen scale of time]. On the other hand, by the same reasons as in the case presented in Fig. 1(a), the curves $\boldsymbol{r}^s(t)$ and $\boldsymbol{r}^s_{mf}(t)$, and $\boldsymbol{r}^h(t)$ and $\boldsymbol{r}^h_{mf}(t)$ are intersected at $t = t_{in}$ ($t_{in} \approx 0.505$ s and $6.597 \times 10^{-3}$ s for the square and hexagonal lattice, respectively).

In conclusion, we have developed a method to numerically simulate the magnetic relaxation in 2D ensembles of dipolar interacting nanoparticles subjected to a bias magnetic field. Within its framework and within the mean-field approximation we have calculated the relaxation law for different ensembles of Co nanoparticles, and we have analyzed its dependence on the correlations of the magnetic moments, bias field and the lattice structure.

[a] Electronic mail: denisov@ssu.sumy.ua

[b] Electronic mail: trohidou@ims.demokritos.gr

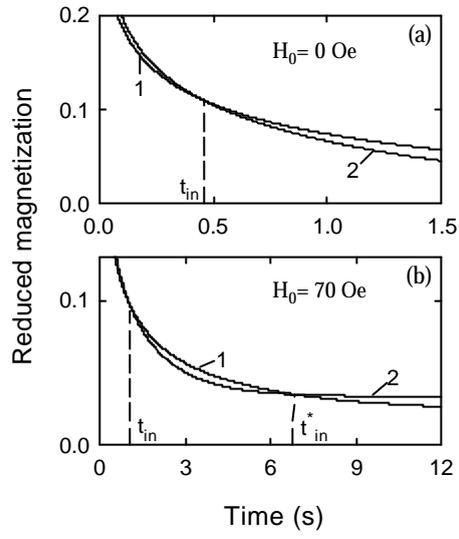

Fig. 1. Plots of $\mathbf{r}^s(t)$ (curve 1) and $\mathbf{r}^s_{mf}(t)$ (curve 2) for $d^s = 3r$.

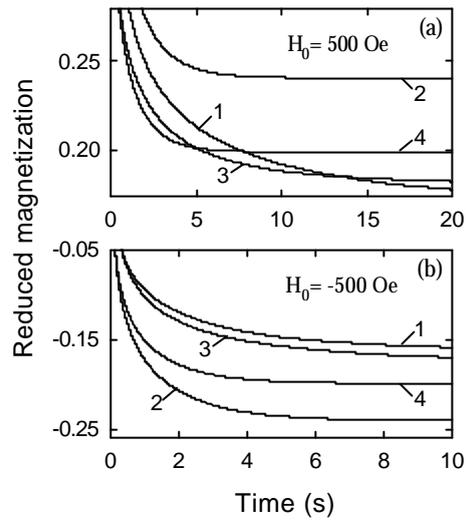

Fig. 2. Plots of $\mathbf{r}^s(t)$ (curve 1), $\mathbf{r}^s_{mf}(t)$ (curve 2), $\mathbf{r}^h(t)$ (curve 3), and $\mathbf{r}^h_{mf}(t)$ (curve 4) for $d^s = d^h = 3r$.

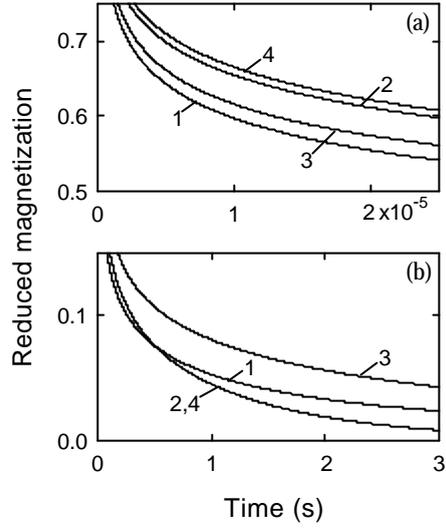

Fig. 3. Plots of $\boldsymbol{r}^s(t)$ (curves 1), $\boldsymbol{r}^s_{mf}(t)$ (curves 2), $\boldsymbol{r}^h(t)$ (curves 3), and $\boldsymbol{r}^h_{mf}(t)$ (curves 4) on small (a) and large (b) time scales for $d^h = 3r$, $d^s = d^h 3^{1/4}/2^{1/2}$, and for $H_0 = 0$.